\documentclass[12pt]{article}

\usepackage{graphicx}
\usepackage{amssymb}
\usepackage[authoryear]{natbib}

\usepackage{siunitx}
\DeclareSIUnit\molar{M}

 \title{Fluctuation spectra of embryonic cell–cell interfaces reveal inverse-square scaling }

\author{Brian Huynh, Shinuo Weng, José Alvarado}
\date{April 2025}

\begin{document}

\maketitle

\begin{abstract}
Tissue-scale shape changes are driven by ensembles of intracellular forces.
However measuring force in these contexts remains a difficult challenge.
Here we perform spectral analysis of transverse fluctuations of cell-cell junctions in \emph{Xenopus} embryonic tissue explants undergoing convergent extension.
We developed an image analysis pipeline to extract fluctuation amplitude profiles $u(x,t)$ from time-lapse confocal movies and computed two-dimensional spatiotemporal power spectra.
We observe power-law scaling of mean-squared fluctuation power spectra consistent with $\langle u_q^2 \rangle \sim q^{-2}$ and $\langle u_f^2 \rangle \sim f^{-2}$.
The spatial scaling agrees with predictions from the Helfrich Hamiltonian, and the temporal scaling agrees with overdamped dynamics of a fluctuating membrane, both in the tension-dominated regime.
Pharmacological reduction of actomyosin contractility (via low-dose blebbistatin or latrunculin B) did not significantly alter either scaling exponent.
Our results provide an early empirical characterization of junction fluctuation spectra in an actively shape-changing tissue.
Simple tension-dominated membrane models appear sufficient to describe transverse junction dynamics despite their active and coupled nature.
This work establishes a quantitative baseline for future studies of tension-bearing tissues and motivates the development of physical models specific to multicellular systems.

\end{abstract}

\section{Introduction}

Cell shape changes are essential for proper biological function.
The actomyosin cytoskeleton is responsible for the contractile forces that underlie many shape change processes.
When placed in collectives such as tissues and embryos, nearby contractile cells experience adhesion forces that transmit actomyosin contractility across cells, giving rise to complex collective cell dynamics. Regulatory processes orchestrate contractility, producing coordinated forces and deformations that drive essential mechanical functions such as embryogenesis and tension homeostasis. How forces at cell scales support mechanical function at tissue scales remains an area of active research.


Although forces are integral to mechanical processes in tissues, measuring or inferring them remains an experimental challenge. Existing methods have made significant progress. Vertex models estimate forces along epithelial tissue, but are highly constrained by the parameters of their energy functions and by the approximation of cells as polygons \cite{fletcher_2014}.
Oil and ferrofluid drop experiments have measured cell forces in tissue but require delicate probing, calibration, and fluid injection \cite{campas_2014,serwane_2017, cheikh_2025a}.
Laser ablation has been used to infer forces from the relaxation times of disrupted tissue, but the method is inherently invasive \cite{zulueta-coarasa_2015}.
These limitations motivate the search for non-invasive, image-based approaches to quantifying mechanical state in living tissues.

Fluctuation analysis offers one such avenue.
Models based on the seminal Helfrich Hamiltonian describes transverse membrane undulations as a balance between thermal forces, surface tension $\sigma$, and bending stiffness $\kappa$ \cite{Helfrich-1973-Z.FurNaturforschungC}.
These models have guided quantitative measurements of tension in red blood cells \cite{brochard_1975,popescu_2006,Turlier-2016-NaturePhys}, giant unilamellar vesicles \cite{Engelhardt-1985-J.Phys.Lett.,Faucon-1989-J.Phys.France,Duwe-1990-J.Phys.France}, giant plasma membrane vesicles \cite{Steinkuhler-2019-CommunBiol}, and, more recently, in vesicles containing active cytoskeletal systems \cite{sciortino_2025a}.
Extending fluctuation analysis from isolated cells and vesicles to cell collectives such as tissues could result in a quantitative and non-invasive force probe in developing embryos.
Experiments on MDCK-based confluent epithelial tissues tracked vertex fluctuations, revealing Rho-dependent dynamics \cite{Fodor-2018-BiophysicalJournal}.
Subsequent theoretical models of confluent epithelial tissues predict longitudinal fluctuations, oscillations, and collapse of cell junctions \cite{Zankoc-2020-BiophysicalJournal}, enabling fluidized regions of tissue \cite{Graham-2025-arXiv.org}.
Recent advances in segmentation-free spectral decomposition of whole-embryo shape change events have identified principal spatiotemporal deformation modes in ascidians \cite{Dokmegang-2025-eLife}.
Additionally, a prior study tracking  subcellular heterogeneities at cell junctions in \textit{Xenopus} tissues identified interplays between cell adhesion molecules and planar cell polarity signaling proteins \cite{Huebner-2021-eLife}. 
Despite these studies, however, spectral analysis of transverse fluctuations in tissue junctions appears to be underexplored.
This kind of study would offer valuable, quantitative information about the forces that underlie cell- and tissue-scale deformations.

Here we perform spectral analysis of transverse fluctuations of cell junctions in a tissue undergoing active morphogenesis.
We extend our prior study of \textit{Xenopus} tissue explants undergoing convergent extension by characterizing the spatial and temporal power spectra of transverse junction fluctuations $u(x,t)$.
We find power-law scalings $\langle u_q^2 \rangle \sim q^{-2}$ and $\langle u_f^2 \rangle \sim f^{-2}$ across our accessible frequency ranges, and show that these scalings persist under pharmacological perturbation of tension. These results provide an early empirical characterization of junction fluctuation spectra in a shape-changing embryonic tissue, establishing a quantitative foundation for theoretical modeling of active junction mechanics. They also suggest that transverse fluctuation analysis may ultimately serve as a non-invasive, imaging-based tool for inferring mechanical properties of junctions in living embryos.

\section{Results}

\begin{figure}
    \centering
    \includegraphics[width=0.9\linewidth]{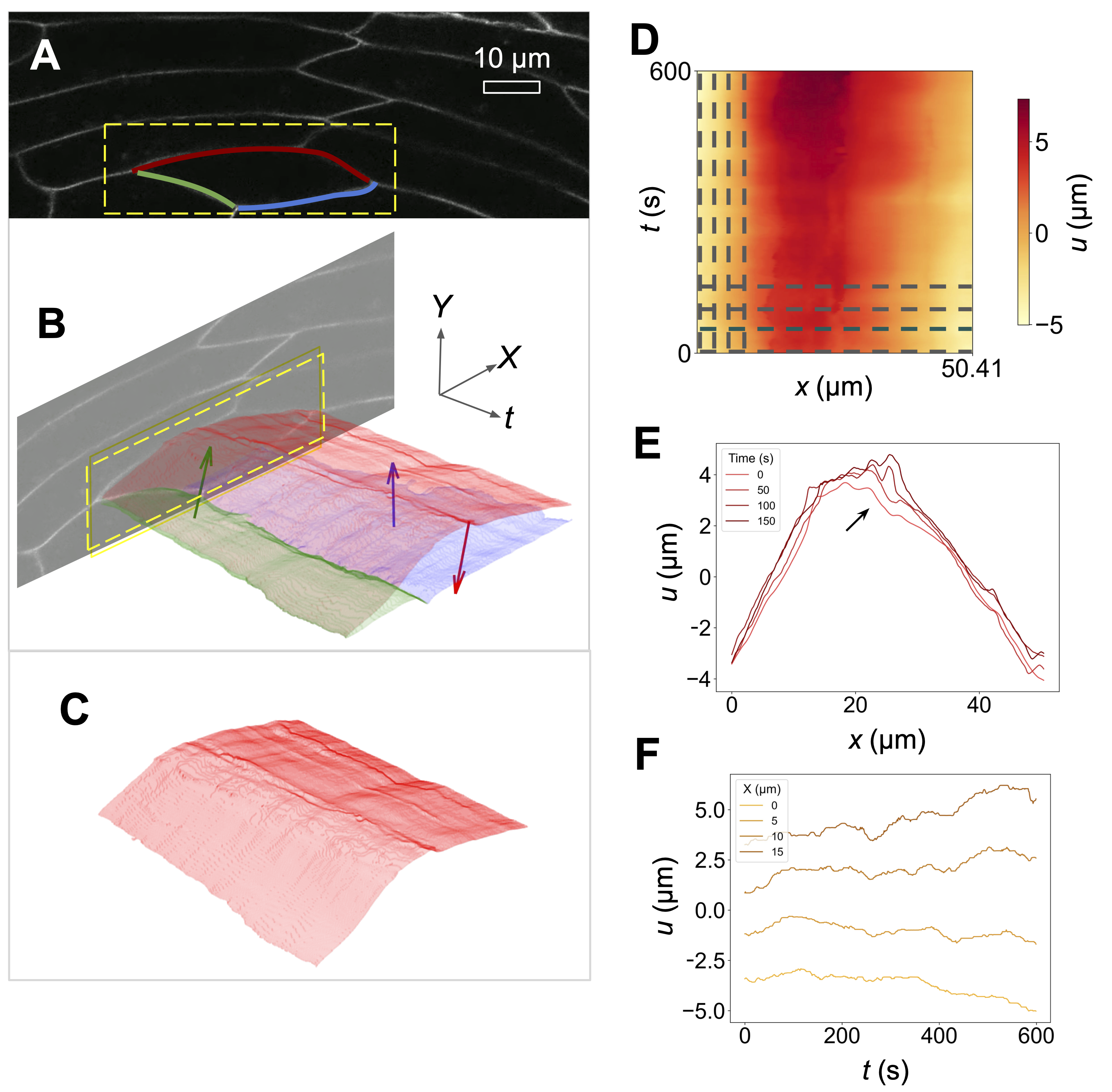}
    \caption{Cell segmentation and Fourier analysis reveals an inverse-square scaling law in the fluctuations in space and time. \textbf{(A)} Representative image in the $(X,Y)$ plane of cells in a \textit{Xenopus} tissue undergoing convergent extension, with the cell membrane labeled (white). Yellow dashed box highlights three selected junctions (red, green, and blue lines). \textbf{(B)} $XYt$-representation of junction fluctuations, with the image in (A) displayed in the $t=0$ plane, and the three selected junctions extruded along the $t$-axis. Arrows denote surface normal vectors evaluated at the base of the vector. \textbf{(C)} One of the selected junctions from (B), isolated to highlight fluctuations. \textbf{(D)} Same junction as in (C), with fluctuation amplitude $u$ expressed as a function of the distance $x$ across the midline and time $t$. The magnitude of $u(x,t)$ is shown as a heat map (color bar, right). \textbf{(E)} Selected rows (D, horizontal gray dashed lines). Black arrow denotes a fluctuation “hotspot” (Discussion). \textbf{(F)} Selected columns (D, vertical gray dashed lines).}
    \label{fig:junctions}
\end{figure}

Our previous study investigated time-lapse movies of \textit{Xenopus} embryo explants undergoing convergent extension \cite{wengPCPdependentPolarizedMechanics2025} (Fig. \ref{fig:junctions}A).
We found that junctions in wild-type tissues undergo transverse fluctuations $|u(x)|$ on the order of $\qty{200}{\nano\meter}$.
Upon downregulation of actomyosin-generated tension, the magnitude of fluctuations increased by an additional $\qtyrange[range-units = single]{30}{50}{\nano\meter}$.
These results showed that tension largely determines the observed transverse fluctuations of cell-cell junctions.
To determine the spectral properties of junction fluctuations, we turn to analysis in Fourier space.

\subsection{Junction transverse fluctuation power spectra scale as $q^{-2}$ and $f^{-2}$}

\begin{figure}
    \centering
    \includegraphics[width=0.5\linewidth]{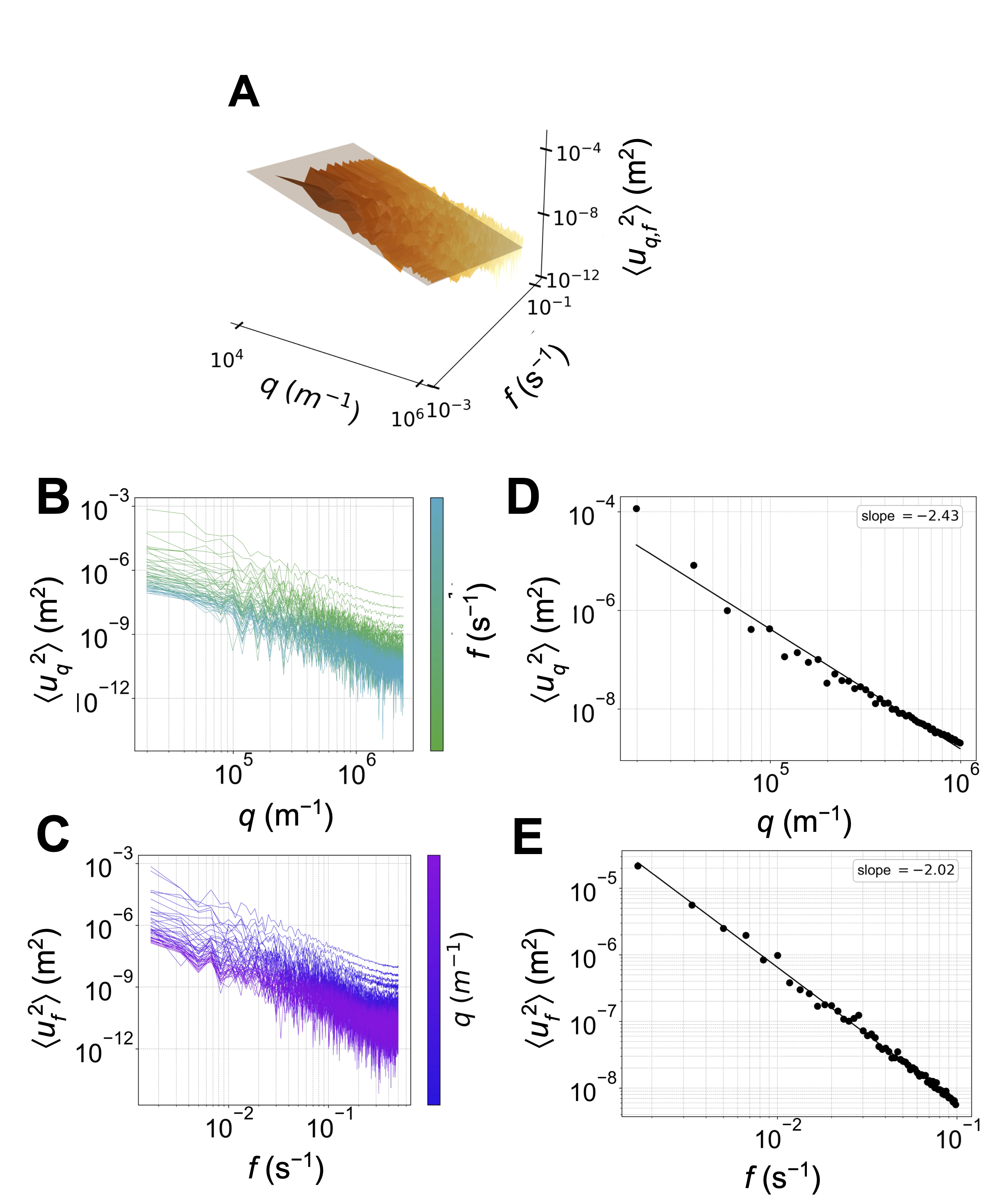}
    \caption{\textbf{(A)} 2D Fourier transform of Fig.~\ref{fig:junctions}D. Color (yellow to red) indicates magnitude. Gray plane denotes best-fit plane. \textbf{(B)} Same data as in (A), with $q$ as abscissa and color indicating $f$ (color bar, right). \textbf{(C)} Same data as in (A), with $f$ as abscissa and color indicating $q$ (color bar, right). \textbf{(D)} Same data in (B), averaged across different values of $f$. Black line denotes best fit of log-transformed data, with slope $-2.43$. \textbf{(E)} Same data in (B), averaged across different values of $q$. Black line denotes best fit of log-transformed data, with slope $-2.02$.}
    \label{fig:fourier}
\end{figure}

To determine spectral properties of junction fluctuations, we develop an image analysis algorithm (Methods).
We capture movies of elongating tissues (pixels in the $(X,Y)$ plane, evolving in time $t$) and track junctions and their fluctuations (Fig. \ref{fig:junctions}B–F).
Next, we determine fluctuation amplitudes $u(x,t)$, with $x$ the distance along the junction's midline.
We perform a discrete spatiotemporal Fourier transform (Python 'numpy.fft.fft2') and square to recover the spectrum $u_{q,f}^2$ (Fig. \ref{fig:fourier}A, yellow-red surface).
To investigate the $q$- (and $f$-) dependencies of $u^2_{q,f}$, we first first plot all $u_q^2$ (and respectively $u_f^2$) curves (Fig. \ref{fig:fourier}B,C).
We observe that the selected junction in Fig.~\ref{fig:junctions}C exhibits a fluctuation spectrum that decreases with $q$ and $f$ in a power law form (Fig. \ref{fig:fourier}B,C).
In order to extract the scaling exponent $\alpha$ of the scaling relation $\langle u_q^2 \rangle \sim q^\alpha$ , we average $u^2_{q,f}$ over all measured $f$ and fit a power law Fig. \ref{fig:fourier}D).
We perform a similar procedure to determine the scaling exponent $\phi$ of the scaling relation $\langle u_f^2 \rangle \sim f^\phi$ by averaging $u^2_{q,f}$ over all measured $q$ (Fig. \ref{fig:fourier}E).

\begin{figure}
    \centering
    \includegraphics[width=0.6\linewidth]{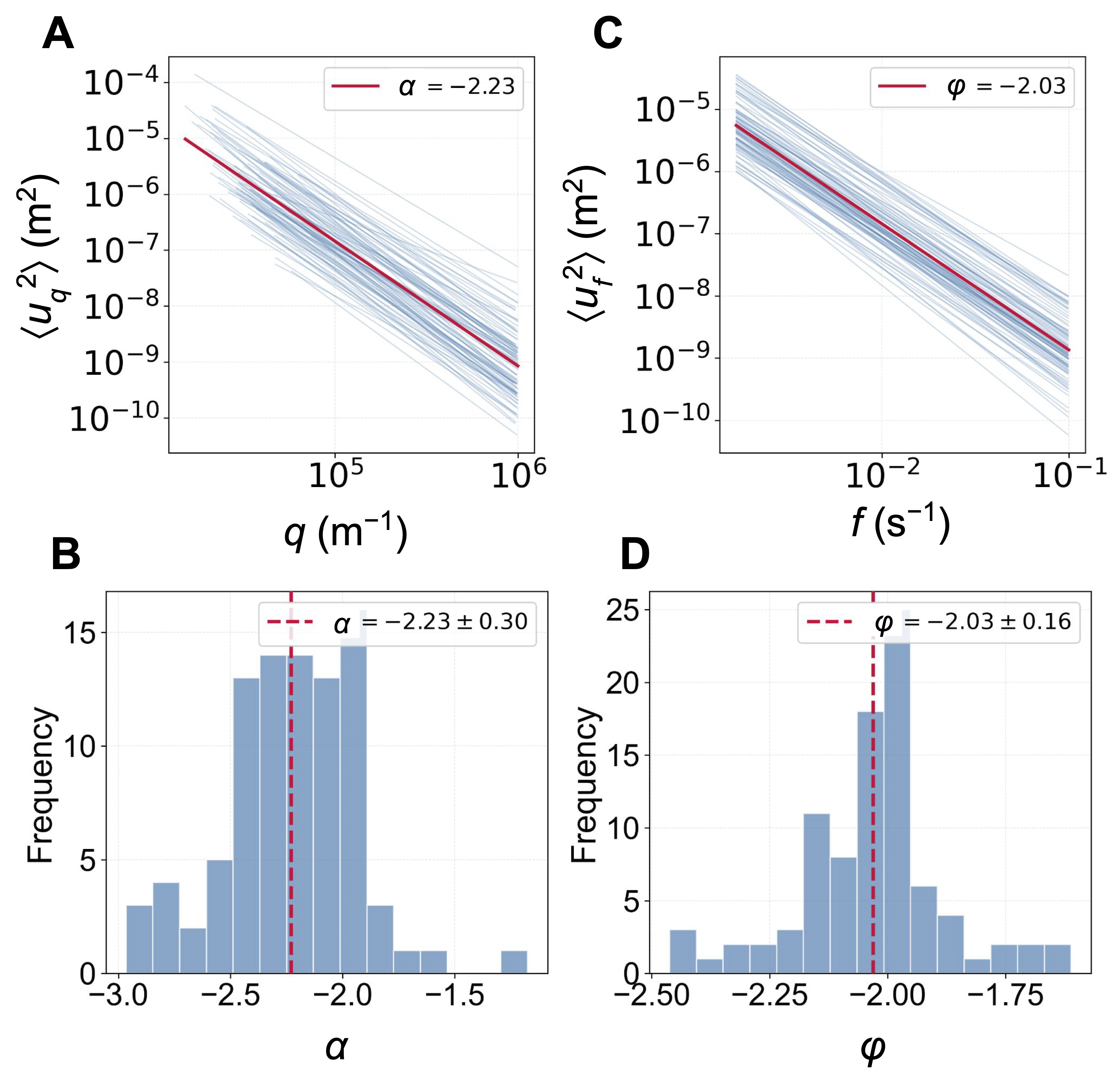}
    \caption{Fluctuation amplitudes exhibit power law scaling consistent with $q^{-2}$ and $f^{-2}$ for wild-type tissues. \textbf{(A)} Plots of best-fit lines of $\langle |u(q)|^2 \rangle_f$ for each individual junction (blue lines). Red line denotes average across all junctions, with slope $\alpha = -2.23$. \textbf{(B)} Probability distribution of best-fit slopes across all junctions. Vertical dashed red line denotes average of $-2.23$. SD of the distribution is $0.30$. \textbf{(C)} Plots of best-fit lines of $\langle |u(f)|^2 \rangle_q$ for each individual junction (blue lines). Red line denotes average across all junctions, with slope $\phi = -2.03$. \textbf{(D)} Probability distribution, as in (B). Vertical dashed red line denotes average of $-2.03$. SD of the distribution is $0.16$.}
    \label{fig:all_wt}
\end{figure}

Fig. \ref{fig:all_wt} shows values of the scaling exponents $\alpha$ and $\phi$ measured for $N=90$ distinct junctions in wild-type tissues.
We observe values of $\alpha$ distributed about $2.23\pm0.30$ (Fig. \ref{fig:all_wt}A,B), which is consistent with the scaling relation $\langle u_q^2 \rangle \sim q^{-2}$.
This scaling relation agrees with the prediction from the Helfrich Hamiltonian for membrane fluctuations in the tension-dominated regime.
Additionally, we observe values of $\phi$ distributed about $2.03\pm0.16$ (Fig.  \ref{fig:all_wt}C,D), which is consistent with the scaling relation $\langle u_f^2 \rangle \sim f^{-2}$.
This scaling relation is consistent with overdamped dynamics of fluctuation modes undergoing tension \cite{brochard_1975}.

\subsection{Pharmocological inhibition of actomyosin contractility maintains $q^{-2}$ and $f^{-2}$ scaling}

Having established that cell junctions under tension demonstrate $q^{-2}$ scaling, we ask whether disrupting tension generation affects this scaling.
In order to test this hypothesis, we turn to two pharmacological interventions.

First, we consider blebbistatin, which inhibits non-muscle myosin ATPase activity and reduces actomyosin contractility.
We acquired movies of \textit{Xenopus} tissue explants ($N=10$ replicates) before treatment, as control measurements.
Next, we treated these tissues with $\qty{20}{\micro\molar}$ blebbistatin diluted in dimethylsulfoxide (DMSO) and Steinberg’s solution.
This amount of blebbistatin is low enough to exclude significant disruption of convergent extension or changes in cell shape or junction movement \cite{devittPCPSeptinsGovern2024,Skoglund-2008-Development}.
We imaged samples and analyzed transverse fluctuation spectra of cell junctions, as before, for before- and after-treatment pairs.

\begin{figure}
    \centering
    \includegraphics[width=0.6\linewidth]{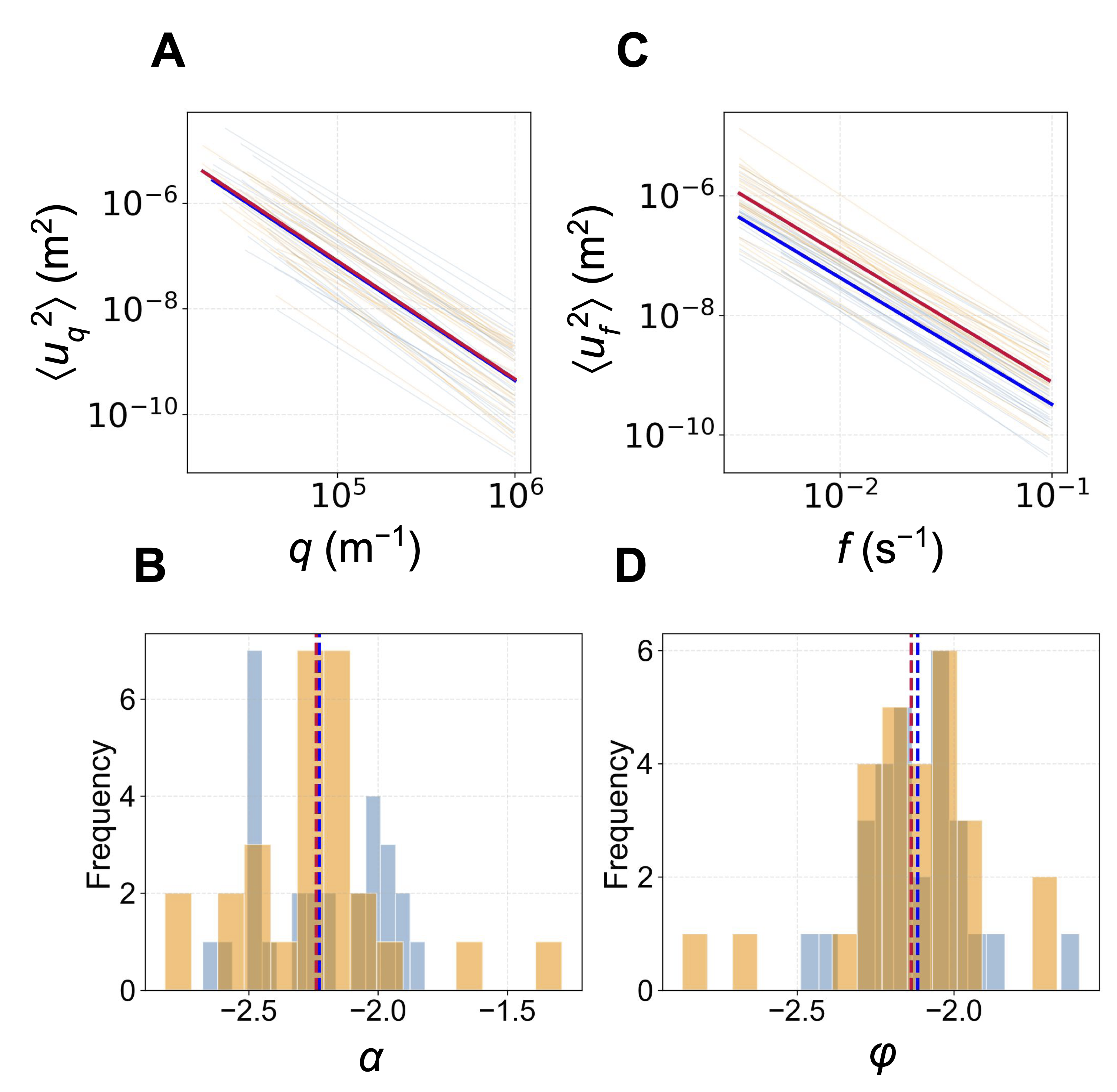}
    \label{fig:blebbistatin}
    \caption{Blebbistatin-treated tissues demonstrate fluctuation scaling relationships consistent with $\langle u_q^2 \rangle \sim q^{-2}$ and $\langle u_f^2 \rangle \sim f^{-2}$. \textbf{(A)} Plots of best-fit lines of $\langle u_q^2 \rangle$ for pre-treatment, control samples (thin blue lines) and post-treatment samples (thin orange lines). Bold lines denote average across all junctions for control samples (blue line, with slope $\alpha = -2.23$) and treated samples (red line, with slope $\alpha = -2.24$). \textbf{(B)} Probability distributions of best-fit slopes across control (blue) and treated (orange) samples. Vertical dashed lines denote the mean values of the control samples (slope of $-2.23$) and the treated samples (slope of $-2.24$). SD of the $\alpha$-distributions are $0.24$ for the control samples and $0.30$ for the treated samples. \textbf{(C)} Plots of $\langle u_f^2 \rangle$, as in (A). Bold lines denote average across all junctions for control samples (blue line, with slope $\phi = -2.12$) and treated samples (red line, with slope $\phi = -2.14$). \textbf{(D)} Probability distributions, as in (B). SD of the $\phi$-distributions are $0.17$ for the control samples and $0.24$ for the treated samples.}
\end{figure}

The mean scaling exponent across junctions is $\alpha = 2.23\pm0.24$ for control samples (Fig. \ref{fig:blebbistatin}A,B; blue lines and bars).
This value is consistent with that found for wild-type samples (2.23±0.30, cf. Fig. \ref{fig:all_wt}A,B).
Blebbistatin treatment did not significantly ($p=0.88$) change the exponent compared to before treatment, with $\alpha = 2.24\pm0.30$ (Fig. \ref{fig:blebbistatin}A,B; orange/red lines and bars).
These results demonstrate that blebbistatin treatment retains $q^{-2}$ scaling.
We additionally find $f^{-2}$ scaling as well.
Control samples resulted in $\phi = 2.12 \pm 0.17$ (Fig. \ref{fig:blebbistatin}C,D; blue lines and bars), which is in relatively good agreement with the wild-type measurement (2.03±0.16, cf. Fig. \ref{fig:all_wt}C,D).
Adding blebbistatin again resulted in an insignificant ($p=0.73$) change, with $\phi = 2.14 \pm 0.24$ (Fig. \ref{fig:blebbistatin}C,D; orange/red lines and bars).

In order to further test the effect of tension down-regulation, we consider latrunculin B, which disrupts F-actin polymerization and reduces stability \cite{kovacs_2004,wakatsuki_2001}.
We again record movies of fluctuating junctions ($N=13$ replicates) before treatment as a control group.
We then applied $\qty{0.5}{\micro\molar}$ Latrunculin B  diluted in DMSO and Steinberg’s solution.
This dose is low enough to not significantly disrupt convergent extension \cite{kimPunctuatedActinContractions2011}.

\begin{figure}
    \centering
    \includegraphics[width=0.5\linewidth]{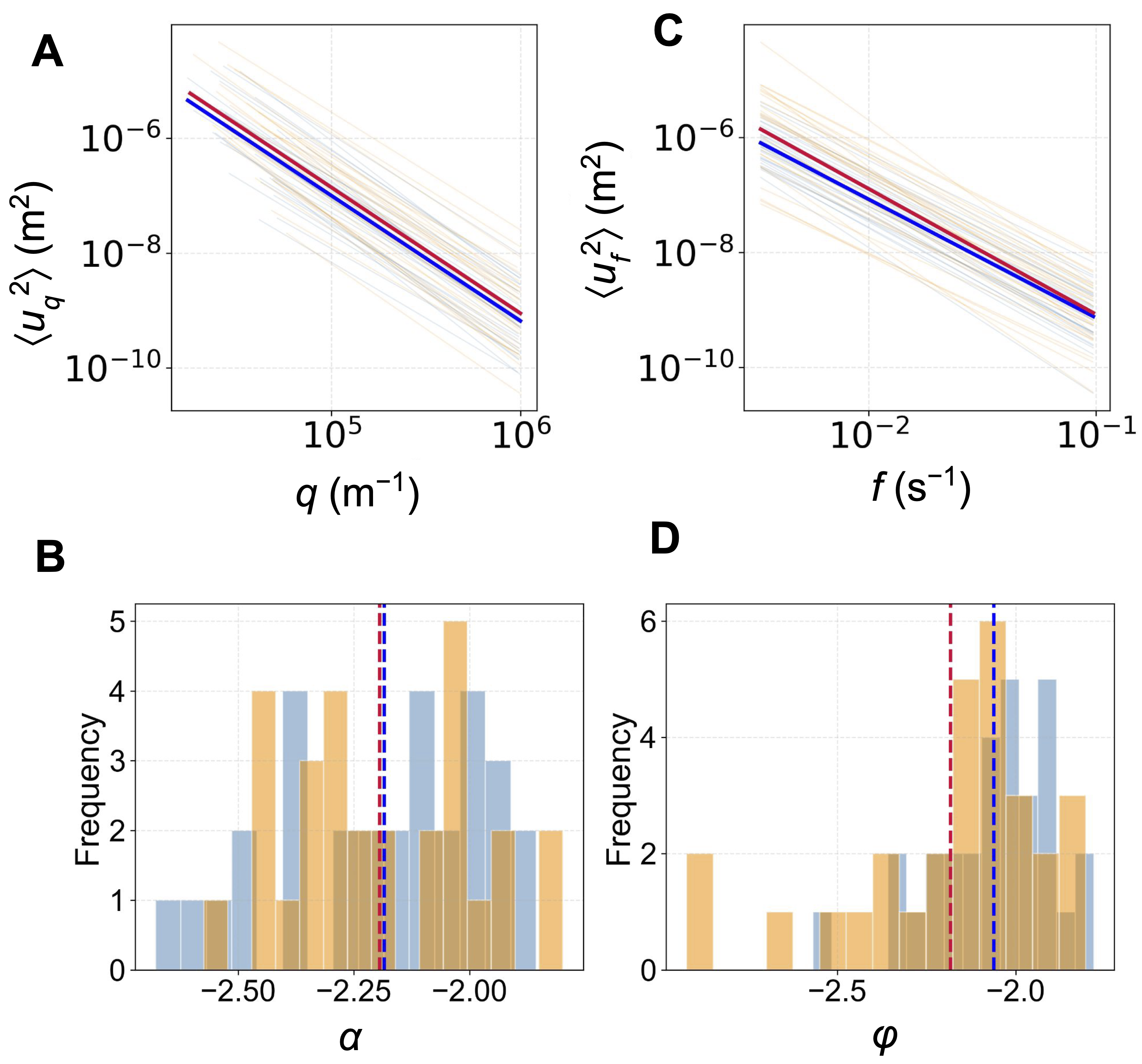}
    \caption{Latrunculin-B-treated tissues demonstrate fluctuation scaling relationships consistent with $\langle u_q^2 \rangle \sim q^{-2}$ and $\langle u_q^2 \rangle \sim f^{-2}$. \textbf{(A)} Plots of best-fit lines of $\langle u_q^2 \rangle$ for junctions in pre-treatment, control samples (thin blue lines) and post-treatment samples (thin orange lines). Bold lines denote average across all junctions for control samples (blue line, with slope $\alpha = -2.18$) and treated samples (red line, with slope $\alpha = -2.19$). \textbf{(B)} Probability distributions of best-fit slopes across control (blue) and treated (orange) samples. Vertical dashed lines denote the mean values of the control samples (slope of $-2.18$) and the treated samples (slope of $-2.19$). SD of the $\alpha$-distributions are $0.22$ for the control samples and $0.20$ for the treated samples. \textbf{(C)} Plots of $\langle u_f^2 \rangle$, as in (A). Bold lines denote average across all junctions for control samples (blue line, with slope $\phi = -2.06$) and treated samples (red line, with slope $\phi = -2.18$). \textbf{(D)} Probability distributions, as in (B). SD of the $\phi$-distributions are $0.17$ for the control samples and $0.27$ for the treated samples.}
    \label{fig:latrunculin}
\end{figure}

We again find that treatment with latrunculin B preserves the $q^{-2}$ and $f^{-2}$ scaling relationships.
Before latrunculin-B treatment, we measure a scaling exponents of $\alpha = -2.18 \pm 0.22$ (Fig. \ref{fig:latrunculin}A,B; blue lines and bars) and $\phi  = -2.06 \pm 0.17$ (Fig. \ref{fig:latrunculin}C,D; blue lines and bars), consistent again with both wild-type (cf. Fig. \ref{fig:all_wt}) and the control group from the blebbistatin experiments (cf. Fig. \ref{fig:blebbistatin}).
Upon Latrunculin-B treatment, we measure $\alpha = -2.19 \pm 0.20$ (Fig. \ref{fig:latrunculin}A,B; orange/red lines and bars) and $\phi = -2.18 \pm 0.27$ (Fig. \ref{fig:latrunculin}C,D; blue lines and bars), with no significant change ($p=0.86, p=0.051$) compared to control.


In sum, our results demonstrate that $q^{-2}$ and $f^{-2}$ scaling relations continue to hold when tension is down-regulated due to mild blebbistatin and latrunculin-B treatment.

\section{Discussion}

We have analyzed the power spectrum of transverse fluctuations of cell junctions in an actively shape-changing tissue.
As a model system, we used \textit{Xenopus} embryonic tissue explants undergoing convergent extension.
We found that junctions exhibited the scaling laws $\langle u_q^2 \rangle \sim q^{-2}$ and $\langle u_f^2 \rangle \sim f^{-2}$.
We observed the same scaling behavior in tissues with reduced tension generation, via treatment with blebbistatin and latrunculin A.
These results agree with the known presence of tension in the tissue, as well as with prior studies investigating how tension affects transverse fluctuations of cell membranes and vesicles.
Overall, our study shows that simple models based on the Helfrich hamiltonian \cite{Helfrich-1973-Z.FurNaturforschungC} and overdamped relaxation dynamics of fluctuation modes \cite{brochard_1975} suffice to describe transverse membrane fluctuations, at least within the range of frequencies studied here.

The frequency ranges we probed resulted from the properties of the confocal microscopy movie acquisitions.
Images were taken with a frame rate of \qty{1.0}{\second^{-1}} and a pixel size of \qty{0.205}{\micro\meter}.
Total movie durations were up to \qty{600}{\second} and junction lengths ranged between \qtyrange{30}{80}{\micro\meter}.
These values resulted in a spatial frequency range of \qtyrange{1.25E4}{4.88E6}{\meter^{-1}} and a temporal frequency range of \qtyrange{1.67E-3}{E-1}{\second^{-1}}.
It is possible that other scaling laws could be observed outside these frequency ranges.
For example, we have not observed a $q^{-4}$ regime, which is expected to occur at higher frequencies.
Observing this regime would be beneficial in developing a method to measure junction tension.
The corner frequency $q_c$, which separates the $q^{-2}$ and $q^{-4}$ regimes, depends on the ratio of tension and stiffness as $q_c = \left(\sigma/\kappa\right)^{1/2}$.
It may possible to observe a $q^{-4}$ dependence elsewhere in Xenopus tissues or in other tissues, particularly in phases of development where the tension is low and/or the inherent bending stiffness of the membrane is high (e.g. due to coupling to the extracellular matrix, or due to increased cell-cell or cell-matrix adhesions).
Measuring $q_c$ would therefore provide a quantitative readout of the tension in a junction.
In addition, it is interesting that we measure an $f^{-2}$ dependency of fluctuation amplitudes to frequencies as low as \qty{E-3}{\second^{-1}}.
A microrheology study of actomyosin active gels found a $f^{-2}$ scaling of the complex part of the response function down to \qty{E-1}{\second^{-1}} \cite{mizunoNonequilibriumMechanicsActive2007d}.
Similar scaling held for red blood cells down to \qtyrange{E-1}{E0}{\second^{-1}} \cite{Turlier-2016-NaturePhys}.
At these low frequencies, myosin motor activity contributed significant, non-equilibrium fluctuations that extended the $f^{-2}$ regime below a corner frequency of \qty{10}{\second^{-1}}, where passive systems would instead exhibit a $f^0$ plateau.

We also observe that fluctuation amplitudes are not always homogeneous, and can contain localized “hotspots” (cf. Fig.~\ref{fig:junctions}E, arrow).
These hotspots could indicate more flexible regions of a junction or an increased bout of localized actomyosin activity.
Our current spectral analysis is not sensitive to these events.
It would be interesting to perform correlation analysis to empirically characterize these spatially non-uniform properties. 

We furthermore observed higher fluctuation magnitudes in tension-downregulated samples compared to control and wild-type.
Comparing the blue and red curves of Fig.~\ref{fig:blebbistatin}A,C, we found that blebbistatin treatment increased the magnitude of $\langle u_f^2 \rangle$. Meanwhile, both $\langle u_q^2 \rangle$ and $\langle u_f^2 \rangle$ increase less upon latrunculin-B treatment.
These observations are consistent with our results from our previous study \cite{wengPCPdependentPolarizedMechanics2025}.
It remains currently unclear which mechanisms allow different drugs to affect $\langle u_q^2 \rangle$ and $\langle u_f^2 \rangle$ in different ways.
However, our observations raise the hypothesis that a closer study of junction fluctuation dynamics could yield rich insights into molecular mechanisms.

Analyzing fluctuations has the potential to provide a non-invasive method to quantitatively measure the tension along a junction.
The only data required is obtained by live imaging, allowing the tissue to remain intact as it develops.
It would be interesting to apply this method to large datasets of developing tissue that track whole-tissue shape changes over multiple phases of development.
In addition to transverse fluctuations, it would be interesting to analyze other related quantities.
Indeed recent studies have investigated vertex displacement correlations and junction longitudinal fluctuations in the context of vertex models \cite{Fodor-2018-BiophysicalJournal, Zankoc-2020-BiophysicalJournal, Graham-2025-arXiv.org}.
Because vertex models often ignore transverse fluctuations, it would be interesting to complement them to provide a more complete picture.
Furthermore, studying the relaxation timescales of individual modes could yield further insight into excitation and relaxation dynamics \cite{sciortino_2025a}.
Overall, our study shows that fluctuation analysis could offer a rich tool to probe tissue mechanics during major shape change events in tissues.

\section{Conclusions}

Existing methods to measure cell membrane tension in tissue are limited and invasive. We anticipate that fluctuation spectroscopy on tissues could provide a non-invasive readout of effective tension along cell junctions. Toward this end, we provide early empirical measurements of the spectral properties of transverse junction fluctuations in \textit{Xenopus} tissue explants undergoing convergent extension.
We observed scaling relations of the power spectral densities of transverse fluctuations consistent with $\langle u_q^2 \rangle \sim q^{-2}$ and $\langle u_f^2 \rangle \sim f^{-2}$.
The $q$ relation is remarkably consistent with simple models of tension-bearing membranes in thermal equilibrium, while the $f$ relation agrees well with overdamped dynamics of a fluctuating line segment.
Future experimental studies could extend the frequency ranges, potentially observing additional scaling laws.

\section{Methods}

\paragraph{Data analysis.}  We begin with triangulated surface meshes of individual cell junctions stored as Wavefront .obj files from our prior study \cite{wengPCPdependentPolarizedMechanics2025}. Each file encodes the three-dimensional geometry of a single cell's membrane surface as a set of vertices with coordinates $(t, X, Y)$ in the laboratory frame — where $X$ and $Y$ are spatial coordinates and $t$ is the time axis of the recording — together with triangular face connectivity and vertex normals. All subsequent analysis was performed using custom Python software.

\paragraph{Junction segmentation.} Because each mesh represents a full cell surface containing multiple junctions, the mesh was first partitioned into $k = 3$ junction regions using k-means clustering on per-face geometric features. For each triangular face, a unit surface normal was computed as the cross product of two edge vectors, and the face centroid was taken as the mean of its three vertex positions. Each face was then represented by a 6-dimensional feature vector formed by concatenating its unit normal (weight 1.0) and its standardized centroid coordinates (weight 0.5). Each feature vector was then normalized by its Euclidean norm prior to clustering. Euclidean k-means was applied to these feature vectors for up to 40 iterations. Following clustering, small isolated fragments — connected components of the face adjacency graph containing fewer than 10,000 faces — were identified by breadth-first search and reassigned to the majority label of their boundary neighbors, ensuring that each junction region formed a single contiguous surface patch. Furthermore, the time-axis component of each cluster centroid was held fixed at its initialized value throughout the iterations, preventing clusters from drifting along the temporal axis. The vertices belonging to each cluster were then collected into a junction object for downstream analysis.

\paragraph{Height-profile extraction.} To extract a transverse displacement profile $u(x, t)$ from each junction, the vertex cloud was reoriented so that the junction midline aligned with the horizontal axis. After reorientation, $x$ denotes the coordinate along the junction midline and $u$ is the transverse displacement of the membrane relative to that midline.
The reoriented vertices were binned into a discrete, two-dimensional height map $u(x, t)$. The number of spatial bins was set so that bin spacing matched the physical pixel size. Each bin was assigned the mean transverse displacement of all vertices falling within it. Empty bins surrounded by a predominantly non-zero $7 \times 7$ neighborhood (at least 40\% non-zero neighbors) were treated as sampling gaps rather than true zeros and were replaced by the mean of their non-zero neighbors within that window. Junctions in which more than 25\% of bins remained empty after this interpolation were considered too sparsely sampled and were excluded from analysis. Finally, to suppress spectral leakage from the spatial boundaries of each junction, 15\% of the grid was cropped from each end of the x-axis, retaining the central 70\% of the junction length for Fourier analysis.

\paragraph{Spectral analysis.} A 2D discrete Fourier transform was applied to the height map $u(x, t)$ to decompose membrane fluctuations into spatial wavenumber $q$ and temporal frequency $f$. The resulting complex amplitudes were zero-centered in both frequency axes and squared to yield the 2D power spectrum $|u(q, f)|^2$. Bins beyond heuristically chosen cutoffs ($\log q > 6$; $\log f > -1$) were excluded from regression, as they were subject to noise.
1D marginal spectra $\langle u_q^2 \rangle_f$ and $\langle u_f^2 \rangle_q$ were obtained by collapsing the 2D power spectrum over the complementary axis, and linear regression was performed on each in log–log space. The resulting best-fit slopes yield the exponents $\alpha$ and $\phi$ for the spatial and temporal spectra, respectively.
Differences in exponent between conditions (DMSO control vs. blebbistatin; DMSO control vs. Latrunculin B) were assessed using Welch's two-sample t-test, applied independently for each slope. Results are reported as mean ± SD with $N$ junctions per condition.

\bibliographystyle{unsrt} 
\bibliography{BBT_TFlux_FFT} 

\end{document}